\documentclass[12pt]{article}
\usepackage{amsmath,amssymb,latexsym,epsfig}
\usepackage{psfrag}
\newcommand{\beq}[1]{\begin{equation}\label{#1}}
\newcommand{\eeq}{\end{equation}}
\def\bse{\boldsymbol{\epsilon}}
\newcommand{\set}[1]{\left\{#1\right\}}

\def\bm{b_{\max}}

\def\cB{{\cal B}}

\def\cA{{\cal A}}

\def\E{{\bf E}}

\def\cQ{{\cal Q}}

\def\d{\delta}
\def\D{\Delta}
\def\e{\epsilon}

\def\G{\Gamma}

\def\n{\nu}

\def\r{\rho}

\def\t{\tau}

\def\Pr{\mbox{{\bf Pr}}}

\def\whp{{\bf whp}}

\newtheorem{theorem}{Theorem}
\newtheorem{corollary}{Corollary}

\newcommand{\brac}[1]{\left( #1\right)}
\newcommand{\bfrac}[2]{\brac{\frac{#1}{#2}}}

\newcommand{\proofend}{\hspace*{\fill}\mbox{$\Box$}}

\begin{document}
\title{{\bf Randomly colouring simple hypergraphs}}
\author{Alan Frieze\thanks{
Supported in part by NSF grant CCF-0502793.}
\ \ \ \   Pall Melsted\\
Department of Mathematical Sciences,\\ Carnegie Mellon University,\\
Pittsburgh PA15213.\\
{\footnotesize Email: alan@random.math.cmu.edu;\ pmelsted@andrew.cmu.edu.}
}
\maketitle
\begin{abstract}
We study the problem of constructing a (near) random proper $q$-colouring of a simple $k$-uniform hypergraph
with $n$ vertices and maximum degree $\D$.
(Proper in that no edge is mono-coloured and simple in that two edges have maximum intersection of size one).
We give conditions on $q,\D$ so that if these conditions are satisfied, Glauber dynamics will converge in $O(n\log n)$
time from a random (improper) start. The interesting thing here is that for $k\geq 3$ we can take $q=o(\D)$.
\end{abstract}

\section{Introduction}
Markov Chain Monte Carlo (MCMC) is
an important tool in sampling from complex distributions. It has been
successfully applied in several areas of Computer Science, most
notably for estimting the volume of a convex body \cite{DFK}, \cite{KLS}, \cite{LV} and estimating the permanent of a non-negative matrix \cite{JSV}. 

Generating a (nearly) random $q$-coloring of 
a $n$-vertex graph $G=(V,E)$ with maximum degree 
$\Delta$ is a well-studied problem in Combinatorics \cite{BW}
and Statistical Physics \cite{SS}.  Jerrum \cite{Jerrum} proved that a simple, popular
Markov chain, known as the {\em Glauber dynamics}, converges to a random
$q$-coloring after $O(n\log{n})$ steps, provided $q/\D>2$.  
This led to the challenging problem of determining the smallest
value of $q/\Delta$ for which a random $q$-coloring can be generated in
time polynomial in $n$. Vigoda \cite{vigoda} gave the first significant improvement over Jerrum's result, 
reducing the lower bound on $k/\D$ to $11/6$ by analyzing a different Markov chain.
There has been no success in extending Vigoda's approach to smaller values of $q/\D$,
and it remains the best bound for general graphs. There are by now several papers giving improvements 
on \cite{vigoda}, but in special cases. See Frieze and Vigoda \cite{FriVig} for a recent survey.

In this paper we consider the related problem of finding a random colouring of a simple $k$-uniform hypergraph.
A $k$-uniform hypergraph $H=(V,E)$ has vertex set $V$ and $E=\set{e_1,e_2,\ldots,e_m}$ are the edges. Each edge is a $k$-subset
of $V$. Hypergraph $H$ is simple if $|e_i\cap e_j|\leq 1$ for $i\neq j$. A colouring of $H$ is proper if every edge contains
two vertices of a different colour. The chromatic number $\chi(H)$ is the smallest number of colours in a proper colouring of $H$.
In the case of graphs $k=2$ we have $\chi(G)\leq \D+1$ but for hypergraphs ($k\geq 3$) we have much smaller bounds. For example
a simple application of the local lemma implies that $\chi(H)=O(\D^{1/(k-1)})$. In fact a recent result of Frieze and Mubayi \cite{FM}
is that for simple hypergraphs $\chi(H)=O((\D/\log\D)^{1/(k-1)})$. The aim of this short paper is to study randomly colouring 
simple hypergraphs when there are fewer than $\D$ colours available.

Before formally stating our theorem we will define the Glauber dynamics.
All of the aforementioned results on colouring graphs (except Vigoda \cite{vigoda})
analyze the Glauber dynamics, which is a simple and popular Markov 
chain for generating a random $q$-coloring.

Let $\cQ$ denote the set of proper $q$-colourings of $H$. For a colouring $X\in\cQ$ we define
$$B_v=B_v(X)=\set{c\in [q]:\;\exists e\ni v\ \text{such that}\ Z(x)=c\ for\ x\in e\setminus\set{v}}$$
be the set of colours unavailable to $v$.

Then let 
$$A(v,X)=Q\setminus B(v,X).$$
For technical purposes, the state space of the Glauber dynamics is $\Omega=Q^V\supseteq\cQ$ where $Q=\{1,2,\dots,q\}$.
From a coloring $X_t\in\Omega$, the evolution $X_t\rightarrow X_{t+1}$
is defined as follows:

{\bf Glauber Dynamics}\vspace{-.2in}
\begin{description}
\item[(a)] Choose $v=v(t)$ uniformly at random from $V$.
\item[(b)] Choose color $c=c(t)$ uniformly at random from $A(v,X_t)$.
\item[(c)] Define $X_{t+1}$ by
$$X_{t+1}(u)=\begin{cases}X_t(u)&u\neq v\\c&u=v\end{cases}$$
\end{description}
We will assume from now on that
\beq{qd}
q\leq 2\D
\eeq
If $q>2\D$ then we defer to Jerrum's result \cite{Jerrum}.

Let $Y$ denote a colouring chosen uniformly at random from $\cQ$. We will prove the following:
\begin{theorem}\label{th1}
Let $H$ be a $k$-uniform simple hypergraph with maximum degree $\D$.
Suppose that \eqref{qd} holds and that for a sufficiently large constant $K$,
\beq{qk}
q^k\geq Kn\D
\eeq
 and that
\beq{Del}
\D\leq \begin{cases}n^{4/3}&k=3\\n^2&k\geq 4\end{cases}.
\eeq
Suppose that the initial colouring $X_0$ is chosen randomly from $\Omega$.
Then 
\beq{rapid}
d_{TV}(X_t,Y)\leq \d\qquad\qquad for\ t\in [t_\d,t^*].
\eeq
where $t^*=e^{q/400k}$ and $t_\d=2n\log(2n/\d)$.

Here $d_{TV}$ denotes variational distance.
\end{theorem}
Note that this theorem only has real content if $(Kn\D)^{1/k}\ll \D$.
The upper bound \eqref{Del} will be needed for an application of the local lemma, see \eqref{lll} below.
Applying the pigeon-hole principle we see that $m\binom{k}{2}\leq \binom{n}{2}$ which implies that $\D\leq n^2$.

Note that we do not claim rapid mixing from an arbitrary start. Indeed, since we are using relatively few colours, it is possible 
to choose an initial colouring from which there is no Glauber move i.e. we do not claim that the chain is ergodic, see Section \ref{blocked}
for examples of blocked colourings.

The theorem has the annoying upper bound of $t^*$ in its formulation. This arises because our coupling argument requires
that a certain condition persists and we can only show that it will persist \whp\ up to a certain time. This means in effect that
we must assume that $\d\geq 2n\exp\set{-\frac{1}{2n}e^{q/400k}}$. On the other hand, this lower bound is very small and it means that
the algorithm will tend to generate colourings that are close to random. Furthermore, it could be used in a standard way, \cite{Jerrum},
to compute an approximation to the number of proper colourings of $H$. 

On the other hand we can prove the following. We can consider Glauber as inducing a graph $\G_{\cQ}$ on $\cQ$ where two colourings are connected
by an edge if there is a move taking one to the other. Note that if Glauber can take $X$ to $Y$ in one step, then it can take 
$Y$ to $X$ in one step.
\begin{corollary}\label{cor1}
The graph $\G_{\cQ}$ contains a giant component $\cQ_0$ of size $(1-o(1))|\cQ|$.
\end{corollary}

\section{Good and bad colourings}
Let $X\in \Omega$ be a colouring of $V$. For a vertex $v$ and $1\leq i\leq k-1$ let
$$E_{v,i,X}=\set{e:\;v\in e\ and\ |\set{X(w):w\in e\setminus\set{v}}|=i}$$
be the set of edges $e$ containing $v$ in which $e\setminus\set{v}$ uses exactly $i$ distinct colours under $X$.
Let $y_{v,i,X}=|E_{v,i,X}|$.

So $|B_v(X)|=y_{v,1,X}$ for all $v,X$ and now let $\bm(X)=\max_{v\in V}y_{v,1,X}$.

Let
$$\e=\frac{1}{8k}.$$

We define a sequence $\bse=\e_1,\e_2,\ldots,\e_{k-2}$ where $\e_1=\e$ and $\e_{i+1}=\e_i/16k$ for $i<k-2$.
We say that $X$ is {\em $\bse$-bad} if $\exists v\in V, 1\leq i\leq k-2$ such that $y_{v,i}\geq \e_i q^i$.
Otherwise we say that $X$ is {\em $\bse$-good}.

In this section we will show that almost all colourings of $\Omega$ are $\bse$-good and almost all colourings in $\cQ$ are $\bse$-good.
Consider a random colouring $X\in\Omega$. We first estimate the probability it is properly coloured. We use the local lemma.

Fix an edge $e\in H$. Then using $\Pr_\Omega$ to indicate the random choice is from $\Omega$,
$$p=\Pr_\Omega(e\ \text{is not properly coloured by}\ X)=\frac{1}{q^{k-1}}.$$
Now consider the dependency graph, in the context of the local lemma. The events are $\cB_e=\set{e\ \text{is not properly coloured})}$.
$\cB_e$ and $\cB_f$ are indpendent if $e\cap f=\emptyset$. Thus the maximum degree $\D_1$ in the dependency graph is bounded
by $k\D$. 
Then
\beq{lll}
4\D_1p\leq 4k\D p\leq \frac{4k\D}{(Kn\D)^{(k-1)/k}}=\frac{4k\D^{1/(k-1)}}{K^{(k-1)/k}n^{(k-1)/k}}<1
\eeq
provided $K^{(k-1)/k}>4k$.

So, by the local lemma, if $m\leq \D n/k$ is the number of edges in $H$,
\beq{bvq}
\Pr_\Omega(X\ \text{is proper})\geq (1-2p)^{m}\geq (1-2p)^{\D n/k}\geq e^{-2p\D n/k(1-2p)}\geq e^{-3\D n/kq^{k-1}}.
\eeq
Given this, we consider the probability that there is a bad vertex.
For a fixed vertex $v$, the value $y_{v,i,X}$ has distribution 
dominated by the binomial $Bin\brac{\D,\binom{k-1}{i}\bfrac{i}{q}^{k-1-i}}$.

So, from Chernoff bound:
\beq{Ch0}
\Pr(B(n,p)\geq \r np)\leq \bfrac{e}{\r}^{\r np}
\eeq
we see that
$$\Pr(y_{v,i,X}\geq \e_i q^i)\leq \bfrac{e\binom{k-1}{i}i^{k-1-i}\D}{\e_i q^{k-1}}^{\e_i q^i}\leq \bfrac{e\D}{\e q^{k-1}}^{\e q}$$
for $i=1,2,\ldots,k-2$.

Now \eqref{qk} and \eqref{Del} imply that
\begin{eqnarray}
(k-1)\log q&\geq& \frac{k-1}{k}\log K+\frac{k-1}{k}\log n+\frac{k-1}{k}\log\D\nonumber\\
&\geq&\frac{k-1}{k}\log K+\frac{1}{k-1}\log\D+\frac{k-1}{k}\log\D\nonumber\\
&=&\frac{k-1}{k}\log K+\brac{1+\frac{1}{k(k-1)}}\log\D.\label{zxc}
\end{eqnarray}
It follows from \eqref{bvq} and \eqref{zxc} that if $X$ is chosen randomly from $\Omega$ then
\begin{eqnarray}
\Pr(X\ is\ \bse-bad)&\leq&kn\bfrac{ek\D}{\e q^{k-1}}^{\e q}\nonumber\\
&=& \exp\set{\log kn-\e q((k-1)\log q-\log\D-1-\log k-\log1/\e)}\nonumber\\
&\leq&\exp\set{\log kn-\frac{\e q}{k(k-1)}\log\D}\qquad\qquad for\ K> (ke/\e)^{k/(k-1)}\nonumber\\
&\leq& \D^{-\e q/(2k(k-1))}.\label{qom}
\end{eqnarray}
So, using $\Pr_\cQ$ to indicate the random choice is from $\cQ$,
\begin{align}
&\Pr_{\cQ}(X\ \text{is }\bse-bad)=\Pr_\Omega(X\text{ is }\bse-bad\mid X\text{ is a proper colouring})\nonumber\\
&\leq kn\bfrac{e\D}{\e q^{k-1}}^{\e q}e^{3n\D/q^{k-1}}\nonumber\\
&\leq \exp\set{\log kn-\frac{\e q}{k(k-1)}\log\D+\frac{3n\D}{q^{k-1}}}\nonumber\\
&\leq \D^{-\e q/(2k(k-1))}.\label{good}
\end{align}
Thus \whp, a random proper or improper $q$-coloring of $H$ is $\bse$-good. 
\section{Persistence of goodness}
We show first that 
\beq{first}
\Pr(X_t\ is\ 2\bse-good\ for\ t\leq t_0\mid X_0\ is\ \bse-good)\geq 1-2^{-\e q/2}.
\eeq
where 
$$t_0=\frac12\min\set{\frac{\e_{k-2}(1-2\e)q^{k-1}n}{ek\D},\frac{qn}{2e}}.$$

For vertices $x,y\in V$ that share an edge, let $e(x,y)$ be that edge.

For a vertex $v$ and $1\leq i\leq k-1$ let $z_{v,i,t}=y_{v,1,t}+y_{v,2,t}+\cdots+y_{v,i,t}$ where $y_{v,i,t}=y_{v,i,X_t}$
for all $v,i,t$. 
Observe that if $X_t$ is $2\bse$-good and $x\in V$ then
\beq{y1}
\Pr(z_{x,i,t+1}-z_{x,i,t}=1\mid X_t)\leq\frac{2ky_{x,i+1,t}}{qn}
\eeq
for $1\leq i\leq k-2$.

We have $z_{x,i,t+1}-z_{x,i,t}=1$ only if (i) $v=v(t)\in E_{x,i+1,X_t}$ and (ii) $X_t(v)$ is a colour used once on $e(v,x)\setminus \set{x}$
and (iii) $c(t)$ is used on $e(v,x)\setminus \set{x,v}$. Now $\Pr((i),(ii),(iii))\leq \frac{2y_{x,i+1,t}}{n}\cdot \frac{k-1}{(1-2\e)q}$. 
(The 2 is only needed for $i=1$ and $k=3$).
This yields \eqref{y1}, since $\e< 1/2k$.

We consider the following sequence of events for $1\leq i\leq k-2$:
$$\cB_i(t)=\set{\exists s\leq t,v\in V:X_\t\text{ is }2\bse-good\text{ for }\t<s\text{ and }z_{v,i,s}\geq
 z_{v,i,0}+\e_i q^i}$$
Let $\cB(t)=\bigcup_{i=1}^{k-2}\cB_i(t)$ and note that if $\neg\cB(t)$ then $X_t$ is $2\bse$-good.

Now $X_0$ is $\bse$-good and so \eqref{y1} implies that so long as $X_\t$ is $2\bse$-good for $\t<s\leq t$, we have
$z_{v,k-2,s}\leq z_{v,k-2,0}+Bin\brac{s,\frac{k\D}{(1-2\e)qn}}$. So, on using \eqref{Ch0},
\beq{y3}
\Pr(\cB_{k-2}(t))\leq tn\bfrac{etk\D/((1-2\e)qn)}{\e_{k-2} q^{k-2}}^{\e_{k-2} q^{k-2}}\leq \frac{t_0n}{2^{\e_{k-2} q^{k-2}}}.
\eeq
The reader will observe that we have not shown that $t_0\geq \e_{k-2}q^{k-2}$. We do not claim this and when $t_0<\e_{k-2}q^{k-2}$ 
we can replace the RHS of \eqref{y3} by zero.

For $i<k-2$ \eqref{y1} implies that so long as $X_\t$ is $2\bse$-good for $\t<s\leq t$ and $\bigcup_{j=i+1}^{k-2}\cB_j(t)$ does not occur, we 
have
$z_{v,i,s}\leq z_{v,i,0}+Bin\brac{s,p}$ where 
$$p=\frac{\e_iq^i+\sum_{j=1}^i\e_jq^j}{(1-2\e)qn}\leq \frac{2\e_iq^{i-1}}{n}.$$ 
So,
\beq{y4}
\Pr\brac{\cB_i(t)\biggr|\neg\bigcup_{j=i+1}^{k-2}\cB_{j}(t)}\leq t_0n\bfrac{2et}{qn}^{\e_i q^i}\leq \frac{t_0n}{2^{\e_i q^i}}.
\eeq
Equation \eqref{first} follows from \eqref{y3} and \eqref{y4}.

We now show that
\beq{step2}
\Pr(X_{t_0}\ is\ \bse-good\mid X_0\text{ is }\bse-good)\geq 1-e^{-2\e q/99}.
\eeq
For this we use the fact that if $X_t$ is $2\bse$-good and $x\in V$ then
\beq{y1a}
\Pr(z_{x,i,t+1}-z_{x,i,t}=-1\mid X_t)\geq\frac{y_{x,i,t}}{n}
\eeq
for $1\leq i\leq k-2$.

We have $z_{x,i,t+1}-z_{x,i,t}=-1$ if (i) $v=v(t)\in E_{x,i,X_t}$, (ii) $X_t(v)$ is used more than once on $e(x,v)\setminus \set{x}$ 
and (iii) $c(t)$ is a colour not used $e(v,x)\setminus \set{x,v}$. 
Now $\Pr((i),(ii),(iii))\geq \frac{2y_{x,i,t}}{n}\cdot \frac{q-k}{q}$. 
This yields \eqref{y1a}.

We couple $y_{x,i,t}$ with a biassed random walk $Y_t,t\geq 0$ on $\set{0,1,2,\ldots,}$. Here $Y_0=\e_iq^i$
and
$$Y_{t+1}=\begin{cases} Y_t+1&Probability\ \ \frac{4k\e_{i+1}q^i}{n}\\
           Y_t-1&Probability\ \ \frac{Y_t}{n}\\
           Y_t&Probability\ \ 1-\frac{4k\e_{i+1}q^i}{n}-\frac{Y_t}{n}
          \end{cases}
$$
If $\cB(t)$ does not occur then $Y_t$ has no lower a chance of increasing by one than $y_{x,i,t}$ and when $Y_t=y_{i,x,t}$ it has no greater
a chance of decreasing by one. We can therefore, conditional on $\neg\cB(t)$, couple $y_{x,i,t},Y_t$ so that $y_{x,i,t}\leq Y_t$ always.
We can therefore prove \eqref{step2} by proving
\beq{step3}
\Pr(Y_{t_0}>\e_iq^i)\leq e^{-\e_iq^i/49}
\eeq
Let $I=\set{t\leq t_0:Y_{t+1}\neq Y_t}$. Then $|I|$ stochastically dominates by $Bin(t_0,4k\e_{i+1}q^i/n)$ and so
\beq{modI}
\Pr(|I|\leq 2t_0k\e_{i+1}q^i/n)\leq e^{-t_0k\e_{i+1}q^i/n}.
\eeq
So assume now that $I=\set{\t_1,\t_2,\ldots,\t_s}$ where $s\geq 2t_0k\e_{i+1}q^i/n$. Let $Z_j=Y_{\t_j}$ for $j=1,2,\ldots,s$.
Then $Z_{j+1}-Z_j=\pm1$ and $\Pr(Z_{j+1}-Z_j=1)\leq 1/3$ if $Z_t\geq \e_iq^i/2$.

Let $j_0=10\e_iq^i$. Then
\beq{j0}
\Pr(Z_j>\e_iq^i/2,\,0\leq j\leq j_0)\leq e^{-\e_iq^i}.
\eeq
This is because to have $Z_j>\e_iq^i/2$ for $0\leq j\leq j_0$ we must have at least $19j_0/40$ +1's in the sequence
$X_{j+1}-X_j,\,0\leq j<j_0$. Assuming this is not the case let $j_1=\max\set{j:Z_j\leq \e_iq^i/2}$.
If $Y_{t_0}\geq \e_iq^i$ then $j_1\leq t_0-\e_iq^i/2$. But, if 
$$\cA_j=\set{Z_j=\e_iq^i/2\text{ and }Z_l>\e_iq^i/2,j<l\leq s}$$ 
then
\beq{j1}
\Pr(\exists j\leq s-\e_iq^i/2:\;\cA_j)\leq t_0e^{-\e_iq^i/48}.
\eeq
This is because if $\cA_j$ occurs then at least one half of the values in the sequence $Z_{j+1}-Z_j$ are +1, whereas the
expected number is at most one third. This completes the proof of \eqref{step3}, and hence \eqref{step2}.
\section{Coupling Argument}
Now consider a pair $X,Y$ of copies of our Glauber chain. Let
$$h(X_t,Y_t)=|\set{v\in V:X_t(v)\neq Y_t(v)}|$$
be the Hamming distance between $X_t,Y_t$. We use describe a simple coupling between the chains and show
that 
\beq{h1}
\E(h(X_{t+1},Y_{t+1})\mid X_t,Y_t)\leq \brac{1-\frac{1}{2n}}h(X_t,Y_t)
\eeq
if $X_t,Y_t$ are both $2\bse$-good.

Our coupling is the same as that used by Jerrum \cite{Jerrum}. The choice of vertex $v(t)$ will be the same in both chains.
We maximally couple the choice of colour in each chain. 
Then, with $v=v(t)$,
$$\Pr(X_{t+1}(v)\neq Y_{t+1}(v)\mid X_t,Y_t)\leq
(k-1)\frac{|E_{v,1,X_t}|+|E_{v,1,Y_t}|}
{(1-2\e)q}.$$
Hence, assuming that $X_t,Y_t$ are both $2\bse$-good for $1\leq t\leq t_0$ we see that
\begin{eqnarray*}
\E(h(X_{t+1},Y_{t+1})\mid X_t,Y_t)&=&\sum_{v\in V}\Pr(X_{t+1}(v)\neq Y_{t+1}(v))\\
&=&\sum_{w\in V}\Pr(v(t)\neq w\ and\ X_{t}(w)\neq Y_{t}(w))\\
&&+\sum_{w\in V}\Pr(v(t)=w\ and\ X_{t+1}(w)\neq Y_{t+1}(w))\\
&=&\frac{n-1}{n}h(X_t,Y_t)+\frac{1}{n}h(X_t,Y_t)\frac{4(k-1)\e}{1-2\e}\\
&\leq&\brac{1-\frac{1}{2n}}h(X_t,Y_t).
\end{eqnarray*}
Summarising, we have shown that with probability at least $1-2\D^{-\e q/(2k(k-1))}$ we have that
both $X_0,Y_0$ are $\bse$-good. If we run the chain for $t_0t^*$ steps then the probaility
that either chain stops being $2\bse$-good is at most $2t_0t^*e^{-2\e q/99}\leq e^{-3\e q/149}$.
Conditional on these events, $\E(h(X_{t_\d},Y_{t_\d})\leq \d/2$ and this implies \eqref{rapid}.
This completes the proof of Theorem \ref{th1}.
\subsection{Proof of Corollary \ref{cor1}}
The proof of Theorem \ref{th1} shows that if $X,Y\in\cQ$ are both $\bse$-good then there is a path from $X$ to $Y$ in $\cQ$
of length $O(n\log n)$. Since almost all of $\cQ$ is $\bse$-good, we are done.
\proofend
\section{Blocked example}\label{blocked}
We choose $m,q$ sufficiently large and we choose a simple $(k-1)$-uniform 
hypergraph $H_1$ with $m$ vertices and $qm$ edges and maximum degree at most $2k^4q$.
The existence of such a hypergraph is easy to show via the probabilistic method. Fix some $0<p<1$ and choose each possible edge
to include with probability $p$. Let $Z_1$ be the number of edges chosen and let $Z_2$ be the number of pairs of edges that share
two or more vertices. We show is that there is a $p$ such that $\E(Z_1-Z_2\mid D)\geq qm$
where $D$ is the event that the maximum degree is at most $2k^4q$. Now $\E(Z_1)=\binom{m}{k-1}p$ and
$\E(Z_2)\leq \binom{m}{k-1}\binom{k-1}{2}\binom{m-2}{k-3}p^2$. Putting $p=\frac{k^4q}{m\binom{k-1}{2}\binom{m-2}{k-3}}$ gives
$\E(Z_1-Z_2)\geq \frac{k^4q(m-1)}{(k-1)^2(k-2)^2}$. Now the degree of a vertex is $Bin\brac{\binom{m-1}{k-2},p}$ which has mean 
$\frac{2k^4q(m-1)}{m(k-1)(k-2)^2}$
and so the probability its degree is greater than $2k^4q$ is exponentially small in $q$. Thus $\E(Z_1-Z_2\mid D)\geq qm$
and our hypergraph exists.

We build a vertex coloured $k$-uniform simple hypergraph $H$ for which the colouring
is proper and for which there are no Glauber moves. We choose disjoint sets $V_1,V_2,\ldots,V_q$ of size $m$ and let $V=V_1\cup V_2\cup\cdots
V_q$. The vertices in $V_i$ are given colour $i$. 
We let $H_i=(V_i,E_i)$ be a copy of the hypergraph $H_1$. Then for each $i$ we define an injective map $f_i$ from $V\setminus V_i\to E_i$. 
This is possible as $|E_i|=qm\geq |V|$. Then for each $x\in V_j$ and $i\neq j$ 
we add the edge $F_{x,i}=\set{x}\cup f_i(x)$
to $H$. These edges block all Glauber moves. Furthermore, we have (i) $F_{x,i}\cap F_{x,i'}=\set{x}$ for $i\neq i'$ and (ii) 
$|F_{x,i}\cap F_{y,i}|\leq 1$ for $x\neq y$ and (iii) $F_{x,i}\cap F_{y,i}=\emptyset$ for $x\neq y,i\neq j$. Thus the hypergraph created is 
simple. Denote the set of edges added so far by $F_1$ and note that $|F_1|\leq q^2m$.

At the moment the degree of a vertex lies in $[q-1,(k^4+1)q-1]$.
We now add random edges $F_2$ 
so that we have more flexibility with the maximum degree. We only consider edges with at most one vertex in each $V_i$
and we add these with probability $\r$. Now let $A_1$ denote the number of pairs of edges in $F_2$ that share two or more vertices and
let $A_2$ denote the number of pairs of edges, one from $F_1$ and one from $F_2$ that share two or more vertices.

Now $\E(|F_2|)=\binom{q}{k}m^k\r$ and we will choose $\r$ so that $\E(|F_2|)\geq 2\E(A_1+A_2)$. Now
\begin{multline*}
\E(A_1+A_2)\leq \\
\binom{q}{k}m^k\binom{k}{2}\binom{q-2}{k-2}m^{k-2}\r^2+q^2m\binom{k}{2}\binom{q-2}{k-2}m^{k-2}\r\leq (qm)^{2k-2}\r^2+k^2q^km^{k-1}\r.
\end{multline*}
This forces us to choose
$$\r\leq \frac{\e_k}{(qm)^{k-2}}$$
for a sufficiently small $\e_k$.

Then we have $\D\in [O(q),\Omega(qm)]$. 
\section{Open Questions}
\begin{description}
\item[(a)] Is it possible to remove the upper bound $t^*$ in Theorem \ref{th1}?
\item[(b)] Can we remove the factor $n$ in \eqref{qk} which comes from our use of the local lemma?
\item[(c)] Can we extend the result to arbitrary $k$-uniform hypergraphs?
\end{description}

\end{document}